\documentclass{elsart}
\usepackage{graphicx}
\usepackage{url}

\begin{document}
\raggedbottom

\begin{frontmatter}

\title{Skin Lesion Diagnosis \\Using Convolutional Neural Networks }

\author[1]{Daniel Alonso Villanueva Nunez}
\author[1]{Yongmin Li}
\address[1]{Brunel University London,
  United Kingdom}

\maketitle

\begin{abstract}
Cancerous skin lesions are the most common malignancy detected in humans, which, if not detected at early stage, may lead to death. Therefore, it is extremely crucial to have access to accurate results at early stages to optimize the chances of survival. Unfortunately, accurate results are only obtained by highly trained dermatologists which are not accessible to most people, especially in low-income and middle- income countries. Artificial Intelligence (AI) seems to tackle this problem because it has proven to provide equal or better diagnosis than healthcare professionals.

This project collects state-of-the-art techniques for the task of image classification from other fields and implements  them  in  this  project.  Some  of  these  techniques  include  mixup,  presizing,  test  time augmentation, among others. These techniques were implemented in three architectures: DenseNet121, VGG16 with batch normalization, and ResNet50. Models are built with two purposes. First to classify images into seven categories – melanocytic nevus, melanoma, benign keratosis like lesions, basal cell carcinoma, actinic keratoses and intraepithelial carcinoma, vascular lesions, and dermatofibroma. Second to classify images into benign or malignant. Models were trained using 8012 images and its performance was evaluated using 2003 images.  This model is trained end-to-end from image to labels and does not require the extraction of hand-crafted features.  
\end{abstract}


\end{frontmatter}

\section{Introduction}
\label{sec:intro}
Skin Cancer is the most common malignancy detected in humans. Some categories of cancerous skin lesions are dangerous because they can potentially spread all over the human skin, and lead to death if not cure in time. One example is melanoma, which is the most dangerous type of skin cancer, and it is expected to appear in 97,920 people in the United States by the end of 2022 according to the U.S. Cancer Statistics Centre \cite{Siegel2022}. 98\% are the chances of a 5-year survival but it reduces to 18\% once it reaches other organs. For all these reasons, it is important to diagnose malignant skin lesions at an early stage, to permit treatment before metastasis occurs. Malignant skin lesions are first diagnosed by visual inspection and potentially continued by dermoscopic analysis, a biopsy and histopathological examination \cite{Esteva2017}.  Dermoscopic  is  a  non-invasive  skin  imaging  approach  which  permits  visualize zoomed  versions  of  the  subdermal  structures  and  surface  of  the  skin.
Non-invasive procedures are always preferred because they do not destroy the lesion and increases the possibility to monitor its evolution \cite{Maarouf2019}. Examination of these images are time-consuming, tedious and require domain-knowledge. Furthermore, the precision and reproducibility of the diagnosis is highly dependent  on  the  dermatologist’s  experience.  Some  studies  show  that  diagnostic´s  precision  on dermoscopic imaging decreases when they are used by inexperienced physicians \cite{binder1995epiluminescence}. To help doctors make more accurate decisions, computer-aided diagnosis (CAD) systems for skin lesions are used \cite{ruffano2018computerassisted}. These CAD systems are able to help doctors make better decisions by detecting lesion borders, removing noise, among others. However, the process of making an appointment to see the doctor for the skin lesion examination, then proceeding with the dermoscopic procedure, and waiting for the CAD system to extract features for assisting the dermatologists on examining the type of lesion, is long. What if there could exist a way to access highly experienced dermatologists from the commodity of your home that can provide highly accurate results by just using your phone?

Artificial intelligence has proven to surpassed human capabilities. For instance, studies from University Hospitals Birmingham shown that AI systems are able to outperform healthcare professionals when it comes to diagnose diseases \cite{liu2019comparison}. Furthermore, it is expected in 2023 that 4.3 billion people will own a smartphone (Alsop, n.d.). Therefore, being able to place these AI systems on people´s phones will immensely benefit millions of people who do not have access to highly trained physicians. In 2017, Esteva et al. reported the first work on the usage of deep learning convolutional neural networks (CNNs) to classify malignant lesions on the skin that could outperform 21 board-certified dermatologists \cite{Esteva2017}. Esteva et al.’s work was innovative because it did not need extensive pre-processing on images such as segmentation or extraction of visual features. This study presents a system, built using state-of-the-art techniques in deep learning, and that does not require the extraction of hand-crafted features, that is able to classify seven types of skin lesions as well as classify lesions into benign and malignant.

The aim of this project is to build a skin lesion image classifier that is able to identify seven types of skin lesions and whether it is benign or malignant, using state-of-the-art techniques and that does not require hand-crafted features. The rest of the paper is organised as follows: A literature review is provided in Section 2. The methods are described in Section 3. Experiments and results analysis are presented in Section 4. Conclusions are drawn in Section 5.

\section{Background}

Artificial intelligence has revolutionized the world of healthcare and medical imaging applications, where various methods such as probabilistic modelling \cite{Kaba:oe2015,Kaba:his2013,Kaba:hiss2014}, graph cut \cite{Salazar:cimi2011,Salazar:his2012,Salazar:jbhi2014,Salazar:icarcv2010,Salazar:jaiscr2012}, level-set \cite{wang2020blood,Wang:jms2015,Wang:icig2015,Wang:jbhi2017,Wang:icig2015_2,dodo2020simultaneous,Dodo:jms2019,Dodo:best2019,Dodo:cbms2017,Dodo:cbms2019,Dodo:access2019,Dodo:bioimaging2018,Dodo:bioimaging2019,Dodo:bioimaging2018_2} have been investigated. Recently, convolutional neural networks (CNN), a technique used in deep learning, have been found to bring the most progress on the automation of medical imagining diagnostic. A plethora of studies have shown that CNN models match or exceed doctors’ diagnostic performance \cite{mcconnell2022integrating,ndipenoch2022simultaneous,ndipenoch2023retinal}.

For  example,  Esteva  et  al.  built  a  model  which  performance  was  on  par  with  21  board-certified dermatologists.  All  skin  images  were  biopsy-proven  and  tested  in  two  binary  classification  cases: keratinocyte carcinoma versus benign seborrheic keratoses; and melanoma versus melanocytic nevi. The model obtained an AUC of 0.96 for carcinoma and an AUC of 0.94 for melanoma. To build this model, Esteva et al. collected images from three places: the Edinburgh Dermofit Library, the Stanford Hospital, and the ISIC Dermoscopic Archive \cite{Esteva2017}. In the test and validation dataset, blurry images and long-distance images were deleted, yet they were kept during training. The InceptionV3 architecture with pre-trained weights from ImageNet was used. During training, a global learning rate equal to 0.001, and a decay factor of 16 every 30 epochs was implemented. RMSProp was the function used to optimize the model. Images were randomly rotated between 0\textdegree  and 359\textdegree. Random cropping and vertically flip with a probability of 0.5 were also used during data augmentation. After augmentation, images increased by a factor of 720.

Haenssle et al. \cite{haenssle2018man} trained a model which performance was compared with 58 international dermatologists. The model was built to classify images into melanocytic nevi or melanoma. The studied shown that the CNN  model obtained  better  performance  than  most  physicians.  The  deep  learning  model  obtained  a specificity of 82.5\%, which is 6.8\% higher than the average obtained by the dermatologists . Google’s InceptionV4 was the architecture used to train this model. Han  et  al.  \cite{han2018classification} used  the  ResNet-152  architecture  to  build  a  model  to  detect  12  different  skin  diseases: dermatofibroma,  pyogenic  granuloma,  melanocytic  nevus,  seborrheic  keratosis,  intraepithelial carcinoma, basal cell carcinoma, wart, hemangioma, lentigo, malignant melanoma, actinic keratosis, and squamous cell carcinoma. The model was compared with 16 dermatologists and obtained an AUC of 0.96 for melanoma. 

Brinker et al. \cite{binker2019convolutional} utilized a pre-trained on ImageNet ResNet50 architecture to train images from the ISIC dataset for a binary classification task to predict melanoma versus nevus. 92.8\% sensitivity and 61.1\% specificity were achieved  as compared to 89.4\% mean sensitivity and 64.4\% mean specificity obtained by 145 board-certified dermatologists from 12 German university hospitals. To train the  model,  two  datasets  were  used:  the  HAM10000  dataset  and  the  public  ISIC  image  archive.  High learning  rates  were  used  at  the  end  of  the  architecture  while  small  learning  rates  were  used  at  the beginning. This technique is called differential learning rates. The scheduler used to train the model was the cosine annealing method. The experiment consisted in 13 epochs. 

Fujisawa et al. \cite{fujisawa2019deep} built a model that outperform 13 board-certified dermatologists.  The model was built to classify skin lesions into malignant and benign. The accuracy of the model was 92.4\%, the sensitivity was 96.3\%  and  the  specificity was 89.9\%. Meanwhile, the  dermatologists obtained  an  overall  accuracy  of 85.3\%. The architecture chosen was GoogLeNet using pre-trained weights. The dataset utilized belongs to the University of Tsukuba Hospital. Images were trimmed to 1000x1000 pixels  from the centre. Each image was randomly rotated by 15 degrees leading and saved as a different figure. This led to an increase of figures by a factor of 24. When the images were passed through the CNN, blur filters and changes in brightness were applied.     

Other studies include that Eltayef et al. \cite{Eltayef:cbms2017,Eltayef:ida2017,eltayef2016detection1,eltayef2016detection2}
reported various methods in combination with  particle swarm optimization and markov random field for lesion segmentation in dermoscopy images.

As it can be seen, there are a plethora of studies that show that deep learning can match or outperform physicians when it comes to the detection of skin lesions. However, not all studies on the usage of CNNs for skin disease detection compared their results with physicians. Nonetheless, they can provide useful information when it comes to the techniques used to train the models.     

Majtner  et  al.  \cite{majtner2018ensemble} achieved  0.801  accuracy  using  VGG16,  0.797\%  accuracy  using  GoogleNet,  and  0.815\% accuracy by assembling both of them. Models were built to classify the same seven types of  skin  lesions  that this  project  is  attempting to  do.  The  ISIC 2018  dataset  was  used. For  both architectures,  only  the  last  layers  were  trained.  To  deal  with  the  unbalanced  data,  each  image  was horizontally flipped, which led dataset to increase by a factor of two. Then, a rotation factor was assigned to each image to balance the data. 

Xie et al. \cite{xie2017melanoma} obtained 94.17\% accuracy using the ISIC 2018 dataset to classify images into benign or malignant which is considered as a binary classification task. The best AUC reported was 0.970. The data to train this model was obtained from the General Hospital of the Air Force of the Chinese People’s Liberation Army. Different methods to extract features from the images were used. For instance, the authors built a self-generating neural network (SGNN) segmentation model to remove hairs from the image. Shape features were not used since many images are not complete. Five colour features were used: RGB features, LUV histogram distances, colour diversity, centroidal distances. Five statistical texture descriptors were computed from a grey level co-occurrence matrix – based texture. They were regional energy, entropy, contrast, correlation, and inverse difference moment. Finally, 6 concavity features were calculated using the lesion convex hull which describes degrees of concavities. To train the model, three artificial  neural  networks  were  ensembled.  The  experiment  run  with  a  max  of  1000  epochs,  sigmoid activation functions were used, and a learning rate of 0.7 was set. Bisla et al. obtained 81.6\% accuracy when using the ISIC 2017 dataset to predict three classes of skin lesion: melanoma,  nevus, and  seborrheic  keratosis  \cite{bisla2019towards}.  A  hair removal  algorithm was implemented for cleaning images. U-Net algorithm was utilized to segment the lesion area from the skin. To tackle the unbalanced problem, generative adversarial networks (GANs) were utilized to generate more images from the labels with less instances. Horizontal flipping, vertical flipping and random cropping were the methods used for data augmentation. The ResNet50 pre-trained on ImageNet architecture was used. 

Almaraz-Damian et al. \cite{almaraz2020melanoma} used transfer learning with different architectures for a binary classification model (benign  vs  malignant)  utilizing  the  HAM10000  dataset with  the  architectures of VGG16,  VGG19,  Mobilenet  v1,  Mobilenet  v2,  ResNet50,  DenseNet-201,  Inception  V3,  Xception respectively. A mix between handcrafted and deep learning features were employed. To balance the data, a SMOTE oversampling technique was implemented. 

Jojoa et al. \cite{Acosta_Caballero_Garcia-Zapirain_Spencer_2021} trained the pre-trained ResNet152 using the dataset for ISIC 2017 to predict skin cancer into benign or malignant. Images were pre-processed by extracting the lesion from the image using Mask R\_CNN. From all the experiments done tweaking the learning rates and number of epochs, the best model obtained had an accuracy of 90.4\%. 

Aljohani et al. \cite{aljohani2022automatic} used transfer learning to predict melanoma or no melanoma on  dataset ISIC 2019 by experimenting different architectures including  DenseNet201,  MobileNetV2,  ResNet50V2, ResNet152V2, Xception, VGG16, VGG19, and GoogleNet. All experiments were run  with  50  epochs,  but  early  stopping  was  used.  To  overcome  the  imbalanced  dataset,  the ImageDataGenerator from TensorFlow was implemented. The best architecture was the GoogleNet with 76\% accuracy with 29 epochs. 

\section{Methods}

In this project, we have selected three different types of network architectures to perform the task of skin lesion recognition: the VGG, ResNet and DenseNet.

\subsection{VGG}

VGG gives reference to a group in Oxford University named Visual Geometry Group \cite{simonyan2015very}. This architecture was built with the aim to understand the relationship between the depth of an architecture and its accuracy. In order to allow the usage of a long structure and avoid using a large quantity of parameters, small 3x3 convolutional kernels were used. At the end of the architecture, three fully connected layers are placed. They are two VGGs normally used: VGG16 which has 16 layers and VGG19 which has 19 layers. VGG16 has 138 million trainable parameters. This project will be using a variant of VGG16 which is VGG16 with batch normalization because it allows faster training and makes the layers’ inputs more stable by re-centring and re-scaling them.    

\subsection{ResNet}

When  deep  networks  are  able  to  commence  converging,  a  degradation  problem  appears.  With  the architecture depth expanding, accuracy saturates leading to fast degradation. Surprisingly, overfitting is not the root of this degradation, and attaching extra layers to an appropriate deep model results to higher training error \cite{he2015deep}. Nevertheless, this problem is tackled by the introduction of deep residual learning framework which is the building block of the ResNet architecture. Instead of passing an input x through a block F(x) and then passing it through a ReLU activation function, the output of F(x) gets added with the input before it gets passed to the next ReLU Activation Function. There are many types of ResNet. Some of its variants are ResNet18, ResNet34, ResNet50, ResNet101, and ResNet152. The number refers to the number of residual blocks stacked.   

This project will be using both ResNet18 and ResNet50. ResNet18 has around 11 million trainable parameters whereas ResNet50 has over 23 million. 

\subsection{DenseNet}

Improving model architecture is not as easy as just stacking new layers because they normally lead to the vanishing  or  exploding  gradient  problem  \cite{glorot2010understanding}.    Nonetheless,  DenseNet  solves  this problem by simplifying the connection between layers which allows the no learning of redundant feature maps. Furthermore, feature reuse is achieved by taking full advantage of this connectivity, this is known as collective knowledge \cite{huang2018densely}. The difference between DenseNet and ResNets is that rather than adding the input  with  the  output  of  the  feature  maps,  they  get  concatenated.  Just  like  ResNet  is  composed  of residuals blocks, DenseNet is composed of dense blocks.  

It  exists  four  variants  of  DenseNet  which  are  DenseNet121,  DenseNet161,  DenseNet169,  and DenseNet201. For this project, DenseNet121 will be used which is composed of 120 convolutions, four average pooling layers, and 8 million trainable parameters.

\section{Experiments}

\subsection{Dataset}

For this project, the HAM10000 dataset is used \\
(\url{https://www.kaggle.com/datasets/kmader/skin-cancer-mnist-ham10000)}.  \\
The dataset consists of 10,015 dermoscopic images belonging to seven categories which can be seen in Table~\ref{tab:skin_lesions}. 

\begin{table}[ht]
\centering
\begin{tabular}{|p{1.5cm}|p{5cm}|p{2cm}|}
\hline
\textbf{Label} & \textbf{Description} & \textbf{Amount} \\
\hline
nv & melanocytic nevi & 6,705 \\
mel & melanoma & 1,113 \\
bkl & benign keratosis – like lesions (solar lentigines / seborrheic keratoses and lichen – planus like keratoses) & 1,099 \\
bcc & basal cell carcinoma & 514 \\
akiec & actinic keratoses and intraepithelial carcinoma / Bowen’s disease & 327 \\
vasc & vascular lesions (angiomas, angiokeratomas, pyogenic granulomas and haemorrhage) & 142 \\
df  & dermatofibroma &  115 \\
\hline
\end{tabular}
\caption{Skin Lesion Labels and Amounts}
\label{tab:skin_lesions}
\end{table}

The  downloaded  file  from  Kaggle  contains  all  the  images  in  two  folders  named “HAM10000\_images\_part\_1”  and  “HAM10000\_images\_part\_2”.  It  also  contains  a  csv  file  named “HAM10000\_metadata.csv” which maps the name of each image file to its label. The csv file has metadata for each image such as the region of the lesion (e.g., ear, head), age of the patient, and sex. An empty folder named “classes” was created containing subfolders with the name of each label.  Then, the images from “HAM10000\_images\_part\_1” and “HAM10000\_images\_part\_2” were map to the subfolders from the folder “classes”. This was done with help of the csv file “HAM10000\_metadata.csv”. This was done such way because that is the format FastAI, which is the framework used for this project, uses.   

The data was split into 80\% training and 20\% validation set which means 8012 images for training and 2003 images for validation. A seed equal to 101096 was used to make this work reproducible. The dataset was only separated in training and validation because some classes have too little images for having a third split for the testing set.  

 \subsection{Data Augmentation}

 Many deep learning practitioners apply data augmentation operations after resizing down the images to a standard form. However, this practice could lead to spurious empty zones, degraded data, or both of them. One example is when an image is rotated 30 degrees and the corners are filled with emptiness which bring not added information to the model. Therefore, to tackle this issue,  presizing  adopts  two  steps.  During  the  first  step,  presizing  resizes  the  images  to  much  larger dimensions  compared  to  the  training  target  dimensions.  This  allows  images  to  have  spare margin  to permit more augmentation transforms on the inner regions without generating empty zones. During the second  step,  it  joins  all  common  augmentation  operations  (including  a  resize  to  the  final  target dimensions) and computes them on the GPU. 

For this work, presizing will be used for data augmentation. First, all images will be expanded to a size of 460x460 pixels in RAM. Then, in the GPU, for each batch, first different augmentations will be stacked to the images. Second, images will be resized to 224x224 pixels and then normalized. That means that for every  batch,  the  computer  is  seeing  a  different  combination  of  transformations.  Nonetheless,  these images  disappear  at  the  end  of  each  batch.  Then,  in  the  next  batch,  a  different  combination  of transformations is applied. This process is performed repeatedly until the training ends. Therefore, if a batch  size  equal  to  64  is  used  to  run  8012  images  for  100  epochs,  then  the  model  will  be  seeing approximately 12,500 images, however, no new images are stored in RAM or in memory.     

\subsection{Results and Analysis}

\begin{figure}[ht]
\begin{center}
\includegraphics[width=\textwidth]{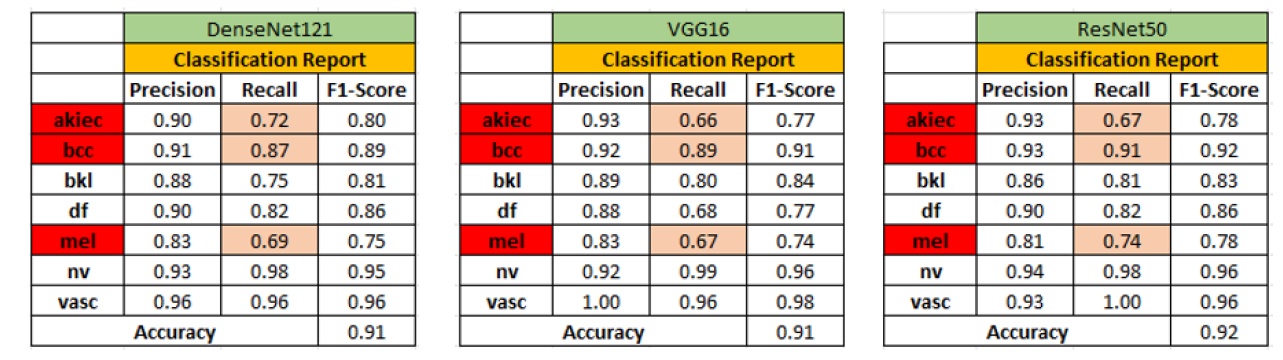}
\end{center}
   \caption{Classification results measured by precision, recall and F1-score for the three network architectures of DenseNet, VGG and ResNet.}
\label{fig:results}
\end{figure}

\begin{figure}[ht]
\begin{center}
\includegraphics[width=0.6\textwidth]{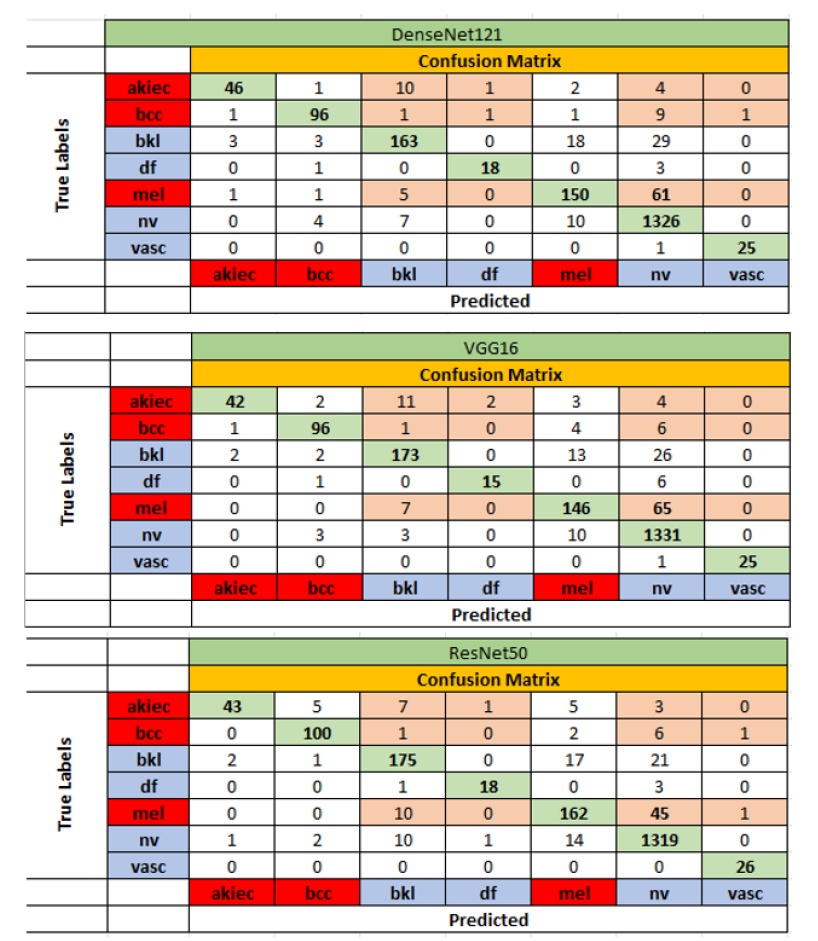}
\end{center}
   \caption{Confusion matrices of all the skin lesion labels for each of the three network architectures of DenseNet, VGG and ResNet.}
\label{fig:confusion_matrix}
\end{figure}

The classification results measured by precision, recall and F1-score for the three network architectures of DenseNet, VGG and ResNet are shown in Figure~\ref{fig:results}. Also, the confusion matrices of all the skin lesion labels for each individual network models are presented in Figure~\ref{fig:confusion_matrix}.

\section{Conclusions}

In this project, we have built skin lesion image classifiers  to identify seven types of skin lesions and whether it is benign or malignant. This task was accomplished by using state-of-the-art techniques in convolutional neural networks for image classification. 
The best model is assessed by the one who obtains the best recalls for the labels akiec, bcc and mel, and the best sensitivity when the model is tweaked for binary classification. The best model found was ResNet50 which achieved a recall for label akiec of 0.67, a recall for label bcc of 0.91, and a recall for label mel of 0.74. When the model was tweaked for binary classification, ResNet50 at a threshold of 0.06 was the one which obtained the best results achieving a sensitivity of 0.9235. In addition, all model’s performances were boosted by using test time augmentation. Here, ResNet50 again obtained the best results with recall 0.69 for label akiec, recall 0.93 for label bcc, and a recall 0.76 for label mel. However, when the model was tweaked for binary classification the model that achieved the best results was VGG16 at threshold 0.05 with a sensitivity of 0.9540.

\bibliographystyle{abbrv}
\bibliography{bib2023}
\end{document}